\documentclass[pra,twocolumn,longbibliography]{revtex4}

\usepackage[utf8]{inputenc}
\usepackage{bm}
\usepackage{mathtools}
\usepackage{bbold}
\usepackage{hyperref}
\usepackage[normalem]{ulem}
\usepackage{color}
\usepackage[dvipsnames]{xcolor}

\begin{document}

\title{Spin squeezing from bilinear spin-spin interactions: two simple theorems}
\author{Tommaso Roscilde, Fabio Mezzacapo and Tommaso Comparin}
\affiliation{Univ Lyon, Ens de Lyon, CNRS, Laboratoire de Physique, F-69342 Lyon, France}


\begin{abstract}
We demonstrate two simple theorems about squeezing induced by bilinear spin-spin interactions that conserve spin parity -- including a vast majority of quantum spin models implemented by state-of-the-art quantum simulators. In particular we show that squeezing captures the first form of quantum correlations which are produced:  1) at equilibrium, by adiabatically turning on the spin-spin interactions starting from a factorized state aligned with an external, arbitrary field; 2) away from equilibrium, by evolving unitarily the same state with the interacting Hamiltonian.   
\end{abstract}
\maketitle


\section{Introduction}
Entanglement \cite{Horodecki2009} is the fundamental trait of quantum many-body states, making them at once very hard to efficiently store in a classical computer; as well as very promising for possible applications in \emph{e.g.} quantum information processing \cite{Nielsenbook} and quantum metrology \cite{Pezze2018RMP}. 
The controlled preparation and certification of entangled quantum many-body states is a central challenge of fundamental research, and it is the pre-requisite for many future quantum technologies \cite{acinetal_roadmap_2018}. 

Among the many forms of entangled states for ensembles of $S=1/2$ spins or qubits, spin-squeezed states \cite{Ma2011PR} occupy a special position, as they are characterized uniquely by the properties of the collective spin variable. Introducing the collective spin operator ${\bm J} = \sum_{i=1}^N {\bm S}_i$ of an ensemble of $N$ qubits, spin squeezed states have the property that the spin squeezing parameter $\xi_R^2 < 1$ \cite{Wineland1994PRA}, where 
\begin{equation}
\xi_R^2 = \frac{ N \min_\perp {\rm Var}(J^\perp)}{|\langle \bm J \rangle|^2}~. 
\end{equation}
Here the minimization is taken over the spin components transverse to the average spin direction $\langle \bm J \rangle$. The presence of spin squeezing $\xi_R^2 < 1$ acts as an entanglement witness, as all separable states have $\xi_R^2 \geq 1$ \cite{Sorensen2001}.  
The metrological use of spin-squeezed states is also straightforward, since spin squeezing allows one to surpass the standard quantum limit for phase estimation \cite{Pezze2018RMP} within Ramsey inteferometry. 

The paradigmatic scheme for the generation of spin-squeezed states via Hamiltonian dynamics is offered by the one-axis twisting (OAT) dynamics \cite{Kitagawa1993PRA}, namely by the evolution of a coherent spin state $|{\rm CSS}\rangle = \otimes_{i=1}^N |\uparrow_z\rangle_i$ (where 
$|\uparrow_z\rangle$ is the $+1/2$ eigenstate of the $S^z$ operator) with the collective-spin Hamiltonian ${\cal H}_{\rm OAT} = \frac{\chi}{N} (J^x)^2$, which is equivalent to an Ising Hamiltonian with infinite-range couplings. Such an Hamiltonian has been realized in seminal experiments on Bose-Einstein condensates \cite{Esteve2008, Riedel2010}; an approximation to the OAT dynamics is offered by long-range Ising interactions realized \emph{e.g.} in ensembles of trapped ions \cite{Bohnet2016}. A second paradigmatic scheme is offered by the two-axis-countertwisting (TACT) dynamics  \cite{Kitagawa1993PRA}, generated by the Hamiltonian ${\cal H}_{\rm TACT} =  \frac{\chi}{iN} \left [ (J^+)^2 - (J^-)^2 \right ] $, which still awaits an experimental realization. 

In this paper we show that spin squeezing is in fact the early form of quantum correlations generated by a large class of bilinear quantum spin Hamiltonians relevant for quantum simulation platforms -- of which the OAT and TACT Hamiltonians and variants thereof are just special cases. The same Hamiltonians can produce spin squeezing either via non-equilibrium dynamics, or via adiabatic dynamics starting from a Hamiltonian which stabilizes the coherent spin state (CSS) as ground state. Our findings apply to bilinear spin Hamiltonians that conserve parity in the computational basis of simultaneous eigenstates of the $S_i^z$ operators, namely that commute with the operator ${\cal P} = \prod_{i=1}^N (2S_i^z)$. Such Hamiltonians take the general form 
\begin{equation}
{\cal H} = {\cal H}_{\perp} + {\cal H}_z  
\end{equation}
where
\begin{eqnarray}
{\cal H}_{\perp} = \sum_{ij} \left (  K_{ij} S_i^+ S_j^+ + J_{ij} S_i^+ S_j^-   + {\rm h.c.} \right)    
\label{e.Ham}
\end{eqnarray}
in which $J_{ij}$ and $K_{ij}$ are generic, complex-valued Hermitian matrices with finite matrix elements. Moreover ${\cal H}_z$ is an arbitrary Hamiltonian term which is diagonal in the computational basis. 

A central role in the discussion will be played by the function
\begin{equation}
{\cal K}(\{\theta_i\})  = \sum_{ij} e^{i(\theta_i+\theta_j)} K_{ij} = R(\{\theta_i\}) + i I(\{\theta_i\})
\label{e.Ktheta} 
\end{equation}
whose real and imaginary parts are given by the real functions $R$ and $I$ respectively. 

 Our findings can be cast in the form of two simple \emph{theorems}, valid for Hamiltonians of the type of Eq.~\eqref{e.Ham}:
\begin{itemize}
\item \emph{Theorem 1}. Consider the initial state  $|\psi(0)\rangle = |{\rm CSS}\rangle$, and the infinitesimally evolved state 
\begin{equation}
|\psi(dt)\rangle = (\mathbb{1} - i {\cal H} dt) |\psi(0)\rangle + {\cal O}(dt)^2~.
\end{equation} 
The state $|\psi(dt)\rangle$ exhibits spin squeezing  iff 
\begin{equation}
I_{\rm max} = \max_{\{\theta_i\}} I(\{\theta_i\})  > 0~,
\label{e.Imax}
\end{equation}  
with squeezing parameter 
\begin{equation}
\xi_R^2[dt] = 1 - \frac{4I_{\max}}{N}~ dt + {\cal O}(dt)^2~.
\end{equation} 

\item \emph{Theorem 2}. Consider the perturbed Hamiltonian 
\begin{equation}
{\cal H}'(\lambda) = \lambda {\cal H} - H \sum_i S_i^z
\end{equation}
 and be $|\psi_0(\lambda)\rangle$ its ground state.
 The state $|\psi_0(d\lambda)\rangle$ exhibits spin squeezing iff
\begin{equation}
R_{\rm max} = \max_{\{\theta_i\}} R(\{\theta_i\})  > 0~,
\label{e.Rmax}
\end{equation}  
 with squeezing parameter
\begin{equation}
\xi_R^2[d\lambda] = 1 - \frac{2R_{\max}}{H N} ~d\lambda + {\cal O}(d\lambda)^2~.
\end{equation} 

\end{itemize}
 
 Recent works \cite{Wangetal2001, Hazzard2013PRL, Hazzard2014PRA, FossFeig2016Arxiv, Perlin2020PRL, Comparinetal2021} have pointed out the robustness of spin-squeezing dynamics in models that deviate from one-axis-twisting Hamiltonians, and that strong squeezing can be found at Ising quantum critical points \cite{Frerot2018PRL_B}. Our theorems include and generalize the Hamiltonians considered in these works.  
 They apply in particular to spin models relevant for quantum simulations, such as the XYZ models with interactions of arbitrary range --  realized using trapped ions \cite{Monroeetal2021}, Rydberg atoms \cite{Browaeys2020NP}, spinful ultracold atoms \cite{Fukuhara2013, Fukuharaetal2015, Mazurenko2017, Jepsen2020}, superconducting circuits \cite{Kjaergaard2020ARCMP}, etc. Therefore all these platforms can act as sources of spin-squeezed states, either adiabatically or dynamically.

\section{Proof of Theorem 1}

It is easy to see that the average collective-spin vector  $\langle \psi(t) | {\bm J} |\psi(t)\rangle$ has (at most) only one finite component, namely the $z$-component $\langle J^z \rangle[t]$. Indeed the bilinear interactions cannot generate any net rotation of the collective spin orientation with respect to the initial state. 

The most general collective spin component transverse to the $z$ axis can be built by defining locally the $x$ and $y$ axes, namely
\begin{eqnarray}
S_i^{x'} & = & \cos\theta_i S_i^x + \sin\theta_i  S_i^y  = \frac{1}{2} \left ( e^{-i\theta_i} S_i^+ + e^{i\theta_i} S_i^{-} \right ) \nonumber \\
S_i^{y'} & = &  -\sin\theta_i S_i^x + \cos\theta_i  S_i^y
\end{eqnarray}
From there, we can define (without loss of generality) $J_\perp(\{ \theta_i \}) = \sum_i S_i^{x'}$. The minimization involved in the definition of the squeezing parameter is then performed with respect to the local angles $\{ \theta_i \}$ .   
 We observe that 
 \begin{eqnarray}
 && \langle \psi(dt) | J^z | \psi(dt)\rangle =  \nonumber \\
 &=& \frac{N}{2} - i \langle {\rm CSS} | [J^z,{\cal H}] | {\rm CSS} \rangle dt + {\cal O} (dt)^2 \nonumber \\
 & = & \frac{N}{2} + {\cal O}(dt)^2
 \end{eqnarray}
 where the linear terms in $dt$ vanish because $|{\rm CSS} \rangle$ is an eigenstate of $J^z$, so that the expectation value of the commutator of $J^z$ with any operator vanishes on that state. 
 
 On the other hand, since $\langle J_\perp \rangle = 0$ at all times, we have that 
 \begin{eqnarray}
 {\rm Var}(J_\perp)[dt] &=& \frac{N}{4} - i \langle {\rm CSS} | [J_\perp^2,{\cal H}] | {\rm CSS} \rangle dt  + {\cal O} (dt)^2~.  
 \end{eqnarray}  
 Moreover we can expand the expression of $J_\perp^2$ as 
 \begin{eqnarray}
J_\perp^2 = \frac{N}{4} &+& \frac{1}{4} \sum_{l\neq m} \Big ( e^{-i(\theta_l +\theta_m)} S_l^+ S_m^+ \nonumber \\
 &&~~~~~~~  + e^{-i(\theta_l -\theta_m)} S_l^+ S_m^- + {\rm h.c.} \Big )
 \end{eqnarray}  
 and take the commutator $[J_\perp^2,{\cal H}]$ piecewise in terms of pairs of spin operators.  When doing so, it is easy to observe that, in the expectation value on the $| {\rm CSS} \rangle$ state, only the latter commutators between products of two spin operators are non zero: 
\begin{eqnarray}
&& \langle {\rm CSS} | [S^+_l S^+_m, S_i^- S_j^-]  | {\rm CSS} \rangle =   \delta_{il} \delta_{jm} + \delta_{im} \delta_{jl} \nonumber \\
 & = & - \langle {\rm CSS} | [S^-_l S^-_m, S_i^+ S_j^+]  | {\rm CSS} \rangle
\end{eqnarray}  
All the other commutators vanish on average over the initial state because of the action of a $S^+$ operator before that of a $S^-$ operator on the same site.
Therefore we can conclude that $\langle {\rm CSS} | [J_\perp^2,{\cal H}] | {\rm CSS} \rangle = - i I(\{\theta_i \})$ where $I$ is defined as in Eq.~\eqref{e.Ktheta}. 
The squeezing parameter (before minimization) takes therefore the form 
 \begin{eqnarray}
 \tilde\xi_R^2[dt] & = & \frac{N^2/4 - NI(\{\theta_i\}) dt + {\cal O}(dt)^2}{N^2/4 + {\cal O}(dt)^2}  \nonumber \\
  & = & 1- \frac{4 I(\{\theta_i\})}{N} dt + {\cal O}(dt)^2~.
 \end{eqnarray}
  Minimizing the expression over the angles $\theta_i$ amounts to maximizing the $I(\{\theta_i\})$ function. Squeezing is therefore present in the state iff $I_{\rm max} = \max_{\{\theta_i \}} I(\{\theta_i\}) > 0$. 
  
  \section{Discussion of theorem 1}
  \label{s.discuss1}
  
  It would be useful to understand for which Hamiltonians of the class of Eq.~\eqref{e.Ham}  the condition of Eq.~\eqref{e.Imax} is satisfied. 
   This class is actually very broad, and it encompasses most of the relevant models for quantum simulators. 
   A most general class of Hamiltonians implemented in the experiments is provided by the XYZ model with arbitrary interactions 
   \begin{eqnarray}
  {\cal H} &=& \sum_{ij}  \left ( {\cal J}^x_{ij} S_i^x S_j^x + {\cal J}^y_{ij} S_i^y S_j^y + {\cal J}^z_{ij} S_i^z S_j^z \right)  \nonumber \\ 
   &=& \sum_{ij}  \left ( \frac{{\cal J}^x_{ij}-{\cal J}^y_{ij}}{4}  ~S_i^+ S_j^+  +  \frac{{\cal J}^x_{ij}+{\cal J}^y_{ij}}{4}  ~S_i^+ S_j^- + {\rm h.c.} \right ) \nonumber  \\
    &+&   \sum_{ij} {\cal J}^z_{ij} ~S_i^z S_j^z 
    \label{e.XYZ}
  \end{eqnarray}
  so that $K_{ij} = ({\cal J}^x_{ij}-{\cal J}^y_{ij})/4 \in \mathbb{R}$. We do not require any spatial symmetry nor condition on the range of the interaction parameters ${\cal J}_{ij}^{\alpha}$ ($\alpha = x,y,z$). 
  
  For this class of models one can easily prove that the condition of Eq.~\eqref{e.Imax}  can be fulfilled provided that 
  \begin{equation}
  R_0 = R(\{\theta_i=0\}) = \sum_{ij} K_{ij} \neq 0~.
  \label{e.K}
  \end{equation}
  This is a very generic condition unless the Hamiltonian is especially fine tuned to have couplings summing up to zero. The condition of Eq.~\eqref{e.K} makes it also clear that the Hamiltonian cannot be $U(1)$ symmetric with symmetry axis along $z$ (in which case one would have ${\cal J}^x_{ij} = {\cal J}^y_{ij}$), but this is obvious from the start, because otherwise the initial state $|{\rm CSS}\rangle$ would be an Hamiltonian eigenstate.  

Indeed, once the condition Eq.~\eqref{e.K} is satisfied, one can choose angles $\theta_i$ such that 
\begin{equation}
\theta_i = \frac{\pi}{4} {\rm sign}(R_0)~~~~ \forall i.
\end{equation}
Since $\exp[i\frac{\pi}{2} {\rm sign}(R_0)] = i ~{\rm sign}(R_0)$, we have that 
\begin{equation}
I \left (\left \{\frac{\pi}{4} {\rm sign}(R_0) \right \} \right) = |R_0| 
\end{equation}
which proves that $I$ can be positive, so that \emph{a fortiori} $I_{\rm max}$ will be positive as well. Even if the condition of Eq.~\eqref{e.K} were not satisfied, it is conceivable that it exists a choice of the angles $\theta_i$ less trivial than the one indicated above, still leading to the conclusion that $I_{\rm max} > 0$.  

Proving the validity of the condition Eq.~\eqref{e.Imax} for generic Hamiltonians, including ones with complex couplings, might be harder; but the theorem ensures that the condition Eq.~\eqref{e.Imax} is necessary and sufficient for squeezing to be generated. For instance, if ${\rm Im}(K_{ij}) \neq 0$ for some pairs $(ij)$, then we can require that 
\begin{equation}
I_0 = I(\{\theta_i=0\}) = \sum_{ij} {\rm Im}(K_{ij}) \neq 0
\label{e.Kpr}
\end{equation}
 -- once again, a generic property away from fine tuning. Under this condition, the uniform choice of angles $\theta_i = 0$ if ${\rm sign}(I_0) = 1$ and $\theta_i = \pi/2$ if ${\rm sign}(I_0) = -1$ guarantees that $I(\{ \theta_i \}) > 0$, so that \emph{a fortiori} $I_{\rm max} > 0$. 

\section{Proof of Theorem 2} 

The state $|\psi_0(d\lambda)\rangle$ can be obtained via first-order perturbation theory \cite{Landau} 
\begin{equation}
|\psi_0(d\lambda)\rangle = |{\rm CSS} \rangle - d\lambda \sum_{n \neq {\rm CSS}} \frac{\langle n | {\cal H} | {\rm CSS} \rangle}{ E_n +H\frac{N}{2}} |n\rangle + {\cal O}(d\lambda)^2
\label{e.psidl}
\end{equation}
where $|n\rangle$ are the eigenstates of the unperturbed Hamiltonian $-H \sum_i S_i^z$ with energies $E_n$. Given that $S_i^+ S_j^- |{\rm CSS}\rangle = S_i^- S_j^+ |{\rm CSS}\rangle = 
S_i^+ S_j^+ |{\rm CSS}\rangle = 0$, we have that only the $S_i^- S_j^-$ in the Hamiltonian Eq.~\eqref{e.Ham} has a non-vanishing action on the CSS, so that 

\begin{equation}
|\psi_0(d\lambda)\rangle = |{\rm CSS} \rangle -  \frac{d\lambda}{2H} \sum_{ij} K^*_{ij} |{\rm CSS}_{ij} \rangle+ {\cal O}(d\lambda)^2
\end{equation}
where $|{\rm CSS}_{ij} \rangle = S_i^- S_j^- |{\rm CSS}\rangle$. 

For this state we have that 
\begin{equation}
\langle J^z \rangle [d\lambda] = \langle \psi_0(d\lambda) | J^z | \psi_0(d\lambda) \rangle   = \frac{N}{2} + {\cal O}(d\lambda)^2 
\end{equation}
since the $J^z$ operator is diagonal in the unperturbed Hamiltonian eigenbasis and the state perturbation in Eq.~\eqref{e.psidl} is orthogonal to the CSS. 

On the other hand, using the same definition of $J_{\rm perp}$ as in Theorem 1, we have that  
$\langle J_\perp \rangle[d\lambda] = 0$
given that $|\psi_0(d\lambda)\rangle$ contains only states $|n\rangle$ with the same parity as that of the CSS, while $J_{\perp}$ is by definition a parity-changing operator. 
Therefore 
\begin{eqnarray}
{\rm Var}(J_\perp)[d\lambda] &=& \frac{N}{4}    -  \frac{d\lambda}{H} ~ {\rm Re} \left (  \sum_{ij} K_{ij}  \langle {\rm CSS}_{ij} | J_\perp^2 | {\rm CSS} \rangle \right ) \nonumber \\
& + &  {\cal O}(d\lambda)^2
\end{eqnarray}
Given that 
\begin{equation}
 \langle {\rm CSS}_{ij} | J_\perp^2 | {\rm CSS} \rangle = \frac{e^{i(\theta_i + \theta_j)}}{2} 
\end{equation}
we conclude that 
\begin{eqnarray}
{\rm Var}(J_\perp)[d\lambda] &=&  \frac{N}{4}    -  \frac{d\lambda}{2H} R(\{ \theta_i \}) + {\cal O}(d\lambda)^2
\end{eqnarray} 
where $R$ is defined as in Eq.~\eqref{e.Ktheta}. Given that the squeezing parameter (before minimization) takes the form 
\begin{equation}
\tilde\xi_R^2 = 1 - \frac{2R(\{ \theta_i\})}{H N}  +  {\cal O}(d\lambda)^2~,
\end{equation}
we can conclude that the perturbed ground state $|\psi_0(d\lambda)\rangle$ features squeezing iff
$R_{\rm max} = \max_{\{ \theta_i \}} R(\{ \theta_i\}) > 0$.

\section{Discussion of theorem 2}
  
Similarly to the discussion of Sec.~\ref{s.discuss1}, we can immediately prove that the theorem applies to the XYZ Hamiltonians of Eq.~\eqref{e.XYZ} under the condition of Eq.~\eqref{e.K} on the Hamiltonian couplings, which turns out to imply the condition of Eq.~\eqref{e.Rmax}. Indeed, choosing $\theta_i = 0$ if ${\rm sign}(R_0) =1$ and $\theta_i = \pi/2$ if ${\rm sign}(R_0) =-1$,  we obtain that $R  = |R_0|>0$, so that $R_{\rm max}$ will be positive \emph{a fortiori}. 
Similarly to what discussed in Sec.~\ref{s.discuss1}, the satisfaction of the condition of Eq.~\eqref{e.K} is sufficient to conclude that the condition of validity of the theorem (Eq.~\eqref{e.Rmax}) is satisfied, but it is not at all necessary -- and it is conceivable to find a combination of angles $\theta_i$ less trivial than the one indicated above, and which still leads to the conclusion that $R_{\rm max} > 0$. 

Similarly to the case of Theorem 1, Theorem 2 can also apply to more general parity-conserving Hamiltonians than those of the XYZ class by exhibiting a set of angles that leads to the satisfaction of the condition Eq.~\eqref{e.Rmax}. 
If ${\rm Im}(K_{ij}) \neq 0$ for some pairs $(ij)$, then we can require the condition Eq.~\eqref{e.Kpr} to be satisfied, which immediately implies that the choice of angles $\theta_i = -\frac{\pi}{4} {\rm sign}(I_0)$ leads to $R = |I_0|$, proving that \emph{a fortiori} $R_{\rm max} > 0$.

\section{Generalization to different initial states}
The above theorems can be straightforwardly generalized to arbitrary initial states $|\sigma_1 \sigma_2, .... \sigma_N\rangle$ in the computational basis,  where $\sigma = +1, -1$ (corresponding to $\uparrow_z, \downarrow_z$ respectively). For that purpose, we can introduce the modified collective spin ${\bm J}' = (J^{x'}, J^{y'}, J^{z'})$ where
\begin{equation}
J^{x'} = \sum_i S_i^{x'} ~~~  J^{y'} = \sum_i \sigma_i S_i^{y'} ~~~ J^{z'} = \sum_i \sigma_i S_i^{z}
\end{equation}

Introducing the unitary transformation $U = \prod_i \exp(i (1-\sigma_i) \pi S_i^x/2)$ we obtain that $U|\sigma_1 \sigma_2, .... \sigma_N\rangle = |{\rm CSS} \rangle$, and $U J^{z'} U^\dagger  = J^z$. Moreover the transformed Hamiltonian ${\cal H}_U = U {\cal H} U^{\dagger}$ maintains the same property of parity conservation -- the only effect of the transformation is that of flipping the signs of some of the couplings. Therefore the theorems apply to squeezing of the modified collective spin ${\bm J}'$ provided that the conditions Eq.~\eqref{e.Imax} and Eq.~\eqref{e.Rmax} apply to the couplings $K_{ij}$ after the sign changes imposed by the unitary transformation. 


\section{Conclusions} 

We have demonstrated via two theorems that, for parity-conserving bilinear Hamiltonians, 
 the first form of entanglement induced by the unitary evolution of a (collinear) coherent spin state of $S=1/2$ spins, or by an adiabatic evolution of the same state induced by the Hamiltonian as a perturbation, is captured by spin squeezing. This result suggests that spin squeezing is an ubiquitous form of quantum correlation, which can be induced by using nearly all current platforms of quantum simulation of quantum spin Hamiltonians. Squeezing appears linearly in time or in the Hamiltonian perturbation, and the rate of buildup of squeezing is uniquely determined by an appropriate sum of the Hamiltonian couplings weighted by phase factors. What happens to the dynamical or adiabatic buildup of squeezing after the initial linear growth depends on further details of the Hamiltonian interactions, such as the number of spatial dimensions in which the Hamiltonian is defined, and the spatial decay of the interactions. These details will dictate whether squeezing -- as quantified by the inverse squeezing parameter $\xi_R^{-2}$ -- grows up to a value which scales with system size (as in the case of the dynamics induced by one-axis-twisting or two-axis-countertwisting Hamiltonians \cite{Kitagawa1993PRA}, or in the case of the adiabatic ground state of Ising Hamiltonians exhibiting quantum-critical points \cite{Frerot2018PRL_B}); or whether it saturates instead to a size-independent value.

\begin{acknowledgments} \emph{Acknowledgements.} We acknowledge useful discussions with B. Laburthe, M. Robert de Saint Vincent and L. Vernac. This work is supported by ANR (``EELS" project) and by QuantERA (``MAQS" project). 
\end{acknowledgments}

\bibliography{refs.bib}

\end{document}